# Electric field variability and classifications of Titan's magnetoplasma environment

C.S. Arridge[1,2], N. Achilleos[3,2], P. Guio[3,2]

1. Mullard Space Science Laboratory, University College London, Dorking, RH5 6NT, UK.
2. The Centre for Planetary Sciences at UCL/Birkbeck, Gower Street, London, WC1E 6BT, UK.
3. Department of Physics and Astronomy, University College London, Gower Street, London, WC1E 6BT.

**Abstract**

The atmosphere of Saturn's largest moon Titan is driven by photochemistry, charged particle precipitation from Saturn's upstream magnetosphere, and presumably by the diffusion of the magnetospheric field into the outer ionosphere, amongst other processes. Ion pickup, controlled by the upstream convection electric field, plays a role in the loss of this atmosphere. The interaction of Titan with Saturn's magnetosphere results in the formation of a flow-induced magnetosphere. The upstream magnetoplasma environment of Titan is a complex and highly variable system and significant quasi-periodic modulations of the plasma in this region of Saturn's magnetosphere have been reported. In this paper we quantitatively investigate the effect of these quasi-periodic modulations on the convection electric field at Titan. We show that the electric field can be significantly perturbed away from the nominal radial orientation inferred from Voyager 1 observations, and demonstrate that upstream categorisation schemes must be used with care when undertaking quantitative studies of Titan's magnetospheric interaction, particularly where assumptions regarding the orientation of the convection electric field are made.

## 1. Introduction

Titan is Saturn's largest moon and the only moon in the solar system known to have a thick atmosphere with an extended exosphere. The formation of Titan's flow-induced magnetosphere is the only known case in the solar system where an essentially unmagnetised body with a thick atmosphere interacts with a magnetospheric plasma. The atmosphere is driven by photochemistry, by charged particle precipitation from the upstream magnetosphere, and presumably by the diffusion of the magnetospheric field into the outer ionosphere. The development of accurate models of the dynamical structure, chemistry, formation and mass loss of Titan's atmosphere rely on an accurate knowledge and characterisation of the local magnetoplasma environment, in order to constrain external sources of energy for the system. See Arridge et al. (submitted manuscruipt, 2011a) for a review of Titan's upstream plasma environment.

The upstream magnetoplasma environment of Titan is a complex and highly variable system and is modulated by internal magnetospheric effects, including apparent longitudinal asymmetries, and external forcing by the solar wind. At Titan's orbital distance (semi-major axis = $1.22 \times 10^6$ km=20.27 $R_S$ where 1 $R_S$=60268 km) the observed magnetic field can become radially stretched at locations just outside Saturn's magnetodisc current sheet, itself a principal external field source (Arridge et al., 2008c). Vertical motions of this magnetodisc (Arridge et al., 2008a; Bertucci et al., 2009; Simon et al., 2010a) can place Titan alternately inside this current sheet or in the lobe-type regions adjacent to the sheet, which are relatively devoid of plasma. The plasma is centrifugally confined in this magnetodisc whereas the energetic particles are more free to extend to higher latitudes (e.g., Achilleos et al., 2010;

Sergis et al., in press). A number of authors have attempted to classify the upstream environment of Titan (Rymer et al., 2009; Simon et al., 2010a; Garnier et al., 2010) using a variety of criteria to produce categories such as "plasma sheet" or "current sheet" or "lobe". The classifications have proven to be useful in understanding flyby-to-flyby variability of Titan's atmosphere (e.g., Westlake et al., 2011). However, such classifications necessarily produce a relatively coarse description of the upstream environment, which does not reduce their value but emphasises that care must be applied in their use, especially for quantitative studies.

Two examples highlight the need for caution. Firstly, energetic neutral atom (ENA) observations can be used to perform remote sensing of Titan's electromagnetic field environment. However, the models used to interpret these observations show that the morphology of the ENA fluxes are highly sensitive to the electric and magnetic field environment of Titan (e.g. Wulms et al. 2010). Hence, making an assumption regarding the electric field orientation and its stationarity based on a categorisation might introduce errors into the interpretation of the ENA observations. As a second example, let us consider the motion of pickup ions produced from Titan's exosphere. Under a "nominal" radial convection electric field orientation pickup ions will be directed away (towards) Titan on the anti-Saturnward (Saturnward) side. However, when the electric field is oriented north-south pickup ion trajectories are in the plane formed by Titan's spin axis and the upstream plasma flow direction. This not only changes the region of velocity space one should search for pickup ions but also can change the location in Titan's atmosphere where pickup ions are deposited thus changing the location of sputtering and heating processes in the thermosphere (e.g, Sittler et al., 2009; Johnson et al., 2009; and references therein). If the electric field magnitude were also to change then this might also increase the magnitude of any atmospheric loss due to pickup.

During the Voyager 1 flyby of Titan the convection electric field was thought to be directed radially away from Saturn because the plasma flows azimuthally past Titan and the background magnetic field was orientated north-south. Cassini studies of the regions upstream of Titan have shown that the magnetic field there is very frequently in a radial orientation (Arridge et al., 2008c; Bertucci et al., 2009; Simon et al., 2010a). It can also be highly variable, changing from a north-south orientation to a radial orientation over timescales of less than one hour (Arridge et al., 2008a; Simon et al., 2010b). These sudden variations in the orientation of the magnetic field can produce significant and rapid rotations in the orientation of the convection electric field in Titan's rest frame.

In this letter we explore the effect of such variability in the upstream magnetic field and quantify the associated variability in the convection electric field. We refer these results to previously established classification schemes and show that classifying Titan's upstream environment as "current-sheet" does not guarantee a particular orientation of the electric field. These results are therefore of relevance in trying to understand the dynamics and evolution of Titan's atmosphere (e.g., Johnson et al., 2009; Sittler et al., 2009) and for interpreting spacecraft data near Titan (e.g., Simon et al., 2007).

## 2. Modelling variations in the convection electric field

We consider the time-dependent variations of the convective electric field **E** in Titan's rest frame. We work throughout in cylindrical coordinates aligned with the current/plasma sheet but assume that the sheet is minimally distorted during its flapping motion such that the unit vectors may be defined as follows: $\mathbf{e}_\rho$ is nearly radially outward from Saturn and lies precisely in the plane of the moving plasma sheet, $\mathbf{e}_z$ is nearly parallel to Saturn's spin axis and orthogonal to the plane of the sheet, and $\mathbf{e}_\varphi$ is parallel to the corotation direction. In this frame the z coordinate of Titan may vary considerably as the current/plasma sheet oscillates about the kronographic equator.

Starting with the definition of **E** and assuming zero radial flow for the magnetospheric plasma one finds the following components of the electric field as measured in the rest frame of Titan:

$$E_\rho = u_z B_\varphi - u_\varphi B_z \qquad E_\varphi = -u_z B_\rho \qquad E_z = u_\varphi B_\rho \qquad (1)$$

In this expression ($E_\rho$, $E_\varphi$, $E_z$) and ($B_\rho$, $B_\varphi$, $B_z$) are the electric and magnetic field components in cylindrical coordinates. The azimuthal component of the plasma velocity, $u_\varphi$ is measured in Titan's reference frame (the plasma speed less Titan's orbital speed of ~5.6 km s$^{-1}$) and $u_z$ is the axial (north-south) speed of the plasma/current sheet. Therefore we recover the pre-Cassini radial electric field ($E_\rho$ = -$u_\varphi B_z$) with a modification introduced by vertical plasma flow in the presence of an azimuthal field component, this vertical plasma flow also introduces an azimuthal electric field when combined with a radial field. The azimuthal plasma flow and radial field then combine to produce the axial electric field. The presence of a radial component in the magnetic field essentially produces a convection electric field which is non-radial – a stronger $B_\rho$ and more rapid azimuthal plasma flow produces a larger axial electric field, and a strong $B_\rho$ and more rapid axial flow produces a larger azimuthal electric field.

One can now ask what the total time derivative of **E** is at Titan as a result of the motion of the plasma sheet and how it might change with the controlling parameters in Saturn's magnetosphere. Figure 1 Illustrates the theoretical setup with three configurations. Consider the plasma and current sheet to be in almost constant axial motion at a speed $u_z$. Because

132 the plane of Titan's orbit is close to Saturn's equatorial plane Titan will appear to move with
133 respect to an observer in the co-moving frame of the plasma sheet. At any given time, Titan
134 may be located at some particular position with respect to this current sheet; it may be located
135 above, in, or below the centre-plane of the sheet. To determine the time derivative of the
136 electric field in Titan's rest frame we need to evaluate the total derivative $D\mathbf{E}/dt = \partial\mathbf{E}/\partial t - u_z d\mathbf{E}/dz$
137 where the negative sign has been introduced to account for the fact that when the plasma
138 sheet is moving up with $u_z > 0$ Titan is actually moving down through the plasma sheet. We
139 assume that the electric field at a fixed point in the sheet to be constant hence $\partial\mathbf{E}/\partial t = 0$ and
140 the total derivative then reduces to the convective derivative associated with the relative
141 motion of the plasma sheet and satellite.
142
143 We write the total derivative $D\mathbf{E}/Dt = -u_z d\mathbf{E}/dz$ and replace the spatial derivative of the
144 azimuthal plasma speed with $(\partial u_\phi/\partial L)(\partial L/\partial z)$ as the product of the equatorial velocity shear of
145 the azimuthal plasma flow, and the stretching of the magnetic field (a more highly stretched
146 field will result in a larger derivative $\partial L/\partial z$) where L is the L-shell of the magnetic field line
147 (defined here as the equatorial radial distance where the magnetic field line crosses the
148 equator). Here we use Ferraro's isorotation theorem (Ferraro, 1937) to write $u_\phi = u_\phi(L)$ where
149 $L = L(\rho, z)$ implying that the plasma on a given field line moves at the same azimuthal velocity.
150 Note that $\partial u_z/\partial z$ is set to zero implying that the axial speed of the plasma sheet does not
151 change across the height of the plasma sheet. Applying this formalism and using these
152 assumptions we find:
153

$$\frac{DE_\rho}{Dt} = u_z \left[ B_z \frac{\partial u_\phi}{\partial L} \frac{\partial L}{\partial z} + u_\phi \frac{\partial B_z}{\partial z} - u_z \frac{\partial B_\varphi}{\partial z} \right]$$

154
$$\frac{DE_\varphi}{Dt} = u_z^2 \frac{\partial B_\rho}{\partial z} \qquad (2)$$

$$\frac{DE_z}{Dt} = -u_z \left[ B_\rho \frac{\partial u_\phi}{\partial L} \frac{\partial L}{\partial z} + u_\phi \frac{\partial B_\rho}{\partial z} \right]$$

155
156 Hence we can see that there is a strong dependence of the time derivative of the convection
157 electric field on how quickly the azimuthal velocity varies with L (velocity shear, $\partial u_\phi/\partial L$), how
158 stretched the field is ($\partial L/\partial z$) and of course how rapidly the plasma sheet is moving in the axial
159 direction ($u_z$). There is also a dependence on the vertical acceleration of the plasma sheet.
160 Hence in a highly stretched magnetodisc with a large velocity shear and intense flapping one
161 can expect rapid changes in the convection electric field strength and direction. We show the
162 azimuthal component for completeness and do not discuss it further since not only is $E_\phi$ the
163 smallest electric field component but its derivative is also the smallest. All of these quantities
164 must be evaluated at the coordinates ($\rho$, $\varphi$, z) of Titan in the plasma sheet coordinate frame.
165

166  To estimate the magnitudes of such temporal modulation of the convection electric field we
167  have used a self-consistent Euler potential model of Saturn's magnetospheric plasma and
168  current sheet (Achilleos et al. 2010). Because this Euler potential model does not contain
169  azimuthal fields we have assumed that $B_\varphi=-0.5B_\rho$ but the exact value of this proportionality
170  does not significantly affect our conclusions. This estimate is obtained from the sweepback
171  angles in Bertucci et al. (2009) and Fig 1. of Arridge et al. (2008a). We couple this with an
172  empirical structural model for the flapping of Saturn's plasma and current sheet (Arridge et al.,
173  submitted manuscript, 2011b).
174
175 $$z_{CS}(t,\varphi,\rho) = \left[\rho - r_H \tanh\left(\frac{\rho}{r_H}\right)\right]\tan\theta_{SUN} + (\rho-\rho_0)\tan\theta_{TILT}\cos\Psi_{PS}(t,\varphi,\rho) \qquad (3)$$
176
177  This "wavy magnetodisc" model (3) gives the vertical position of the plasma sheet as a
178  function of time and radial distance, from which $u_z$ can be easily computed. In this expression,
179  $r_H$, $\theta_{SUN}$, $\rho_0$, and $\theta_{TILT}$ are constants which are obtained by fitting the model to data, and $\Psi_{PS}$ is
180  a phase function which is a function of time (varying between 0 and $2\pi$ over one
181  "magnetospheric period", equal to $2\pi/\Omega$), local time $\varphi$, time t, and radial distance $\rho$.
182
183  Because this is time-dependent it can be differentiated to give the vertical speed $u_z$ of the
184  plasma sheet (4) required to calculate (2). In this derivative the "DC" warping term (Arridge et
185  al.,2008b) from (3) is a constant and becomes zero. The maximum speed of this current
186  sheet is obtained when $\Psi_{PS}=90°$ and maximum acceleration when $\Psi_{PS}=0°$. We use values of
187  $\rho_0=12\ R_S$ and $\theta_{TILT}=12°$ as obtained by Arridge et al. (submitted manuscript, 2011b). Using
188  these values at Titan's orbital distance we find $u_z(\Psi_{PS}=90°)=17$ km s$^{-1}$. In this expression,
189  $\Omega=d\Psi_{PS}/dt$ is the angular frequency of Saturn's global magnetospheric oscillations.
190
191 $$u_z(t,\varphi,\rho) = -(\rho-\rho_0)\Omega\tan\theta_{TILT}\sin\Psi_{PS}(t,\varphi,\rho) \qquad (4)$$
192
193
194  **3.   Results**
195  In figure 2 we present the magnetic field profile through the current sheet from the Euler
196  potential model and the calculated convection electric fields. As expected from (1) only the
197  radial and azimuthal electric field components exhibit any dependence on the speed of the
198  sheet, between maximum and zero speed cases, due to the presence of $u_z$ in these
199  expressions. In each case the effect is fairly modest between the static and moving current
200  sheet cases. Note that these have been plotted for the current sheet moving down over the
201  spacecraft $u_z<0$. For positive current sheet speeds the azimuthal and radial components are 202  reflected about the x axis. This is because the equations for $E_\rho$ and $E_\varphi$ are modulated by $u_z$
203  and swap sign when the plasma sheet is moving in a different direction.

205  The location of the current sheet/lobe boundary has been estimated from the criteria of Simon
206  et al. (2010a) where the current sheet is defined as $B_\rho<=0.6B$. To calculate **B** we have
207  ignored the artificial $B_\varphi$ introduced to solve equations (1) and (2) as this was not calculated
208  self-consistently; doing so results in an artificially large estimate of the current sheet
209  thickness. At the centre of the sheet the radial component dominates whereas at the
210  sheet/lobe boundary (indicated by the vertical dotted lines) the radial and axial electric field
211  components are roughly equal in magnitude, and in the lobes the axial component dominates.

213  In figure 3 the derivatives given by equations (2) are plotted. Because the derivatives are zero
214  when the current sheet is static we plot the case for the sheet moving down over the
215  spacecraft ($u_z<0$, solid line) and moving up over the spacecraft ($u_z>0$, dashed line). The curve
216  for $DE_\rho/Dt$ is not precisely reflected about the x axis because of the $u_z^2$ term in (2). We
217  exclusively consider the $u_z<0$ case (solid line) considering the consequences for Titan moving
218  up through the sheet (also see figure 1c).

220  At the centre of the current sheet near z=0 the electric field is radial and the largest derivative
221  is $DE_z/dt$ consistent with $E_z>0$ above the equator and $E_z<0$ below the equator (odd-
222  symmetry). In the lobes outside of the vertical dotted lines (from the criteria of Simon et al.
223  (2010a)) the electric field derivatives are of a similar magnitude and are approximately 0.1 mV
224  m$^{-1}$ hour$^{-1}$. This is a substantial fraction of the magnitude of the convection electric field
225  showing that large rapid changes in **E** occur over hour-long timescales. It is interesting to note
226  that within the current sheet $E_z$ changes rapidly, reaching a significant fraction of its lobe field
227  value within the current sheet (see figure 2), whereas the largest changes in $E_\rho$ are found
228  near the lobe. Consequently for a spacecraft located in the current sheet with the plasma
229  sheet moving rapidly over the spacecraft, large changes in $E_z$ and rotations of up to 45° in the
230  orientation of **E** will be found before the spacecraft has entered the lobe. This has important
231  consequences for the classification of Titan's upstream environment where although the
232  upstream environment may be jointly classified as current sheet (e.g., Simon et al. 2010a)
233  and plasma sheet (e.g., Rymer et al. 2009), which might lead to assumptions regarding the
234  orientation of the convection electric field, the orientation of this field may be highly non-
235  stationary. In such circumstances the use of these criteria may affect studies which make
236  assumptions about the orientation and magnitude of **E**, for example the interpretation of ENA
237  observations (e.g., Wulms et al. 2010), pickup ion trajectories and subsequent ionospheric
238  heating and loss (e.g. Sittler et al., 2009; Johnson et al. 2009; and references therein), and
239  modelling of the orientation of the induced magnetotail, and modelling ionospheric currents,
240  ionospheric heating and the thermosphere (e.g., Cravens et al. 2009; Ågren et al. 2010).

The total time derivative of $E_z$ is completely dominated by the second term on the right-hand-side, $u_\varphi \partial B_\rho/\partial z$, due to the rapid azimuthal plasma speed and the large vertical gradients in the radial field produced by the current sheet. The shear in the azimuthal plasma velocity ($\partial u_\varphi /\partial L$) was found to be fairly small thus contributing to the small magnitude of the first term. Similarly the total time derivative of the $E_\rho$ is dominated by the $u_\varphi \partial B_z/\partial z$ term even though the vertical gradients in $B_z$ are relatively small. The next important term is the $u_z \partial B_\varphi/\partial z$ term and is smaller than the $u_\varphi \partial B_z/\partial z$ term because the axial speed of the plasma sheet is much smaller than the azimuthal convection speed of Saturn's magnetospheric plasma (17 km s$^{-1}$ Vs. ~120 km s$^{-1}$) and the radial component of the magnetic field is larger than the azimuthal component.

## 4. Conclusions

In this paper we have theoretically examined the effects of magnetodisc flapping on the convection electric field at Titan. We have found that even when classification schemes place Cassini in the current sheet near a Titan flyby, which might indicate a radial electric field, there may in fact be an equally important north-south component. Users of classification schemes must be aware that such modulations might not be captured by these schemes. However, it is important to note that this does not devalue methods for classifying Titan's plasma environment but does highlight the need for caution in their use, particularly for quantitative studies.

The picture of a quasi-static upstream environment where either the spacecraft is in a constant homogeneous lobe or current sheet field configuration does not appear to be valid in the Cassini era (e.g., Arridge et al., 2008b, 2008c; Bertucci et al., 2009; Simon et al., 2010a,2010b; Arridge et al., submitted manuscript, 2011b), possibly as a result of seasonal forcing on Saturn's magnetosphere or particular magnetospheric conditions encountered by Voyager 1. The picture now emerging from various studies is that of a current and plasma sheet in constant motion, with both long (near the planetary rotation period) and short period (due to magnetospheric dynamics driven by internal processes and solar wind variability) motions. This study identifies and classifies the different sources and controlling parameters for the dynamics and variability of the convective electric field.

Further work is required to address the consequences of such rapid changes in the electric field on Titan's induced magnetosphere, ionosphere and atmosphere.

**Acknowledgements**

CSA thanks Nick Sergis and the referee for useful comments on the manuscript. CSA was in this work by a Science and Technology Facilities Council Postdoctoral fellowship (under grant ST/G007462/1), the Europlanet RI project (under grant number 228319), and the International Space Science Institute. CSA thanks staff at the International Space Science Institute for their hospitality during a meeting where this work was first presented. We thank the many individuals at JPL, NASA, ESA and numerous PI and Co-I institutions who have contributed to making the Cassini project an outstanding success.

**Figures**

**Figure 1**

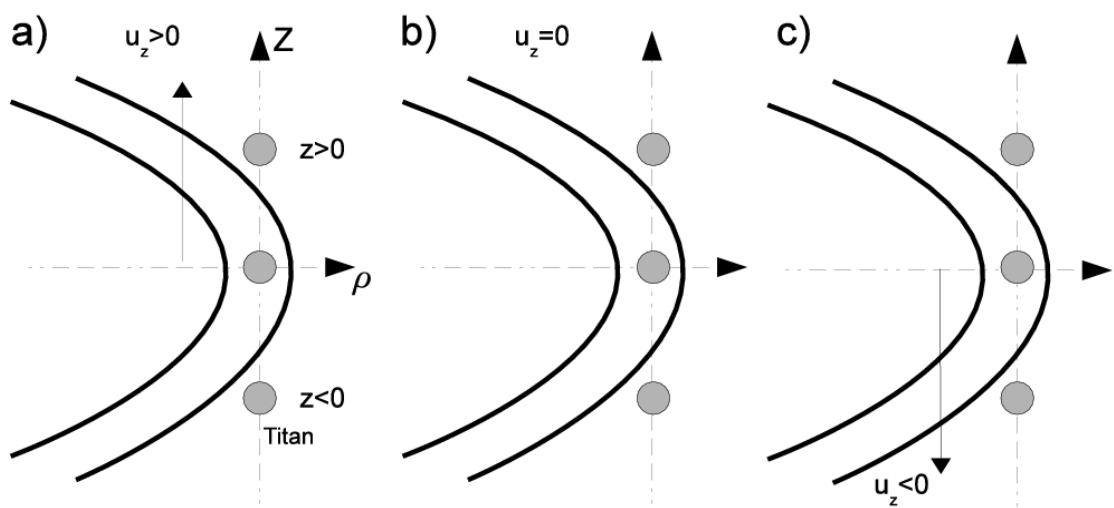

**Figure 1:** Illustration of the theoretical construction. The solid curves indicate magnetic field lines around a current sheet in the z=0 plane. The (z,ρ) coordinates indicate a convenient "sheet coordinate system". Panel (a) shows the sheet moving upwards, (b) the sheet at rest, (c) the sheet moving downwards. In each case Titan (the grey circle) may be located above the sheet, near the centre of the sheet, or below the sheet. For example, the bottom circle in panel (b) would indicate Titan located below the sheet and the sheet at rest with respect to Titan. As a second example, the top circle in panel (a) would indicate Titan located above the sheet but with the sheet moving up to meet Titan, so Titan will shortly pass through the centre of the sheet.

**Figure 2**

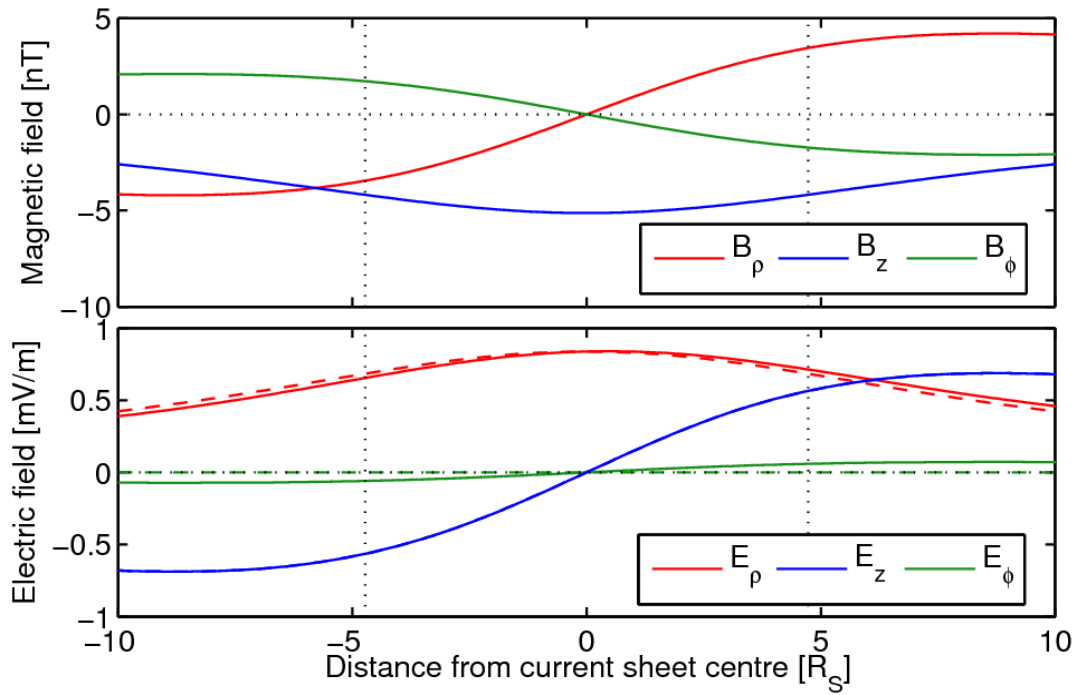

**Figure 2:** Calculated magnetic field profiles (top) and convection electric fields (bottom) as a function of axial distance from the centre of the current sheet, positive above the current sheet and negative below. In both panels red indicates the radial component, blue the axial, and green the azimuthal. The electric field is calculated for two cases i) where the current sheet is in rapid axial motion ($u_z \neq 0$) (solid lines) and ii) where the current sheet is at rest with respect to Titan ($u_z=0$) (dashed lines). The dotted vertical lines indicate the extents of the current sheet as defined by the $B_\rho \leq 0.6B$ criterion of Simon et al. (2010a).

**Figure 3**

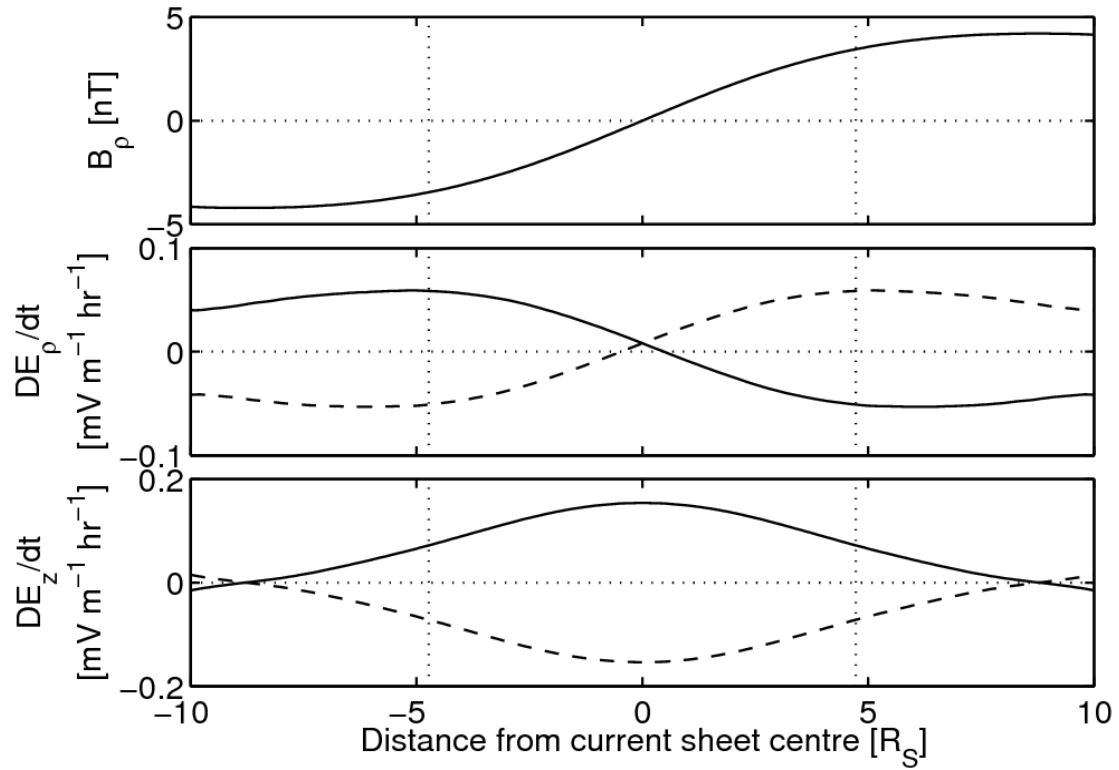

**Figure 3:** Calculated electric field derivatives as a function of axial distance from the centre of the current and plasma sheet. The top panel shows the radial component of the field for reference. The middle and bottom panels show the total time derivative of the radial and axial components of the electric field. The vertical dotted lines show the extents of the current sheet. The solid lines show the rate of change of the electric field when the sheet is in rapid axial motion downwards ($u_z<0$) and the dashed line when the sheet is moving upwards ($u_z>0$).